\documentclass[fleqn,twoside]{article}
	
\usepackage{insa}
\usepackage{graphicx}
\def\lsim{\buildrel{\scriptscriptstyle <}\over{\scriptscriptstyle\sim}}
\def\gsim{\buildrel{\scriptscriptstyle >}\over{\scriptscriptstyle\sim}}
\newcommand{\br}{\begin{eqnarray*}}
\newcommand{\er}{\end{eqnarray*}}
\newcommand{\be}{\begin{equation}}
\newcommand{\ee}{\end{equation}}

\newcommand{\Cpm}{\tilde \chi^\pm}
\def \N0{\tilde\chi^0}
\def \Cpm{\tilde\chi^\pm}

\def \stau1{\tilde\tau_1}
\def \MET {E{\!\!\!/}_T} 

\def \ptau {P_\tau}
\newcommand \pol {{polarization}~}
\newcommand{\taujet}{$\tau$-jet~}
\newcommand{\ch}{H^\pm}
\newcommand{\lumi}{{fb}^{-1}}

\title{
Using Tau Polarization for Charged Higgs Boson and 
SUSY searches at LHC}

\author{Monoranjan Guchait\address{
Department of High Energy Physics,\\
Tata Institute of Fundamental Research,\\
Homi Bhabha Road, Mumbai - 400005, India.
}, D.~P.~Roy.\address{
Homi Bhabha's Centre for Science Foudantion\\
Tata Institute of Fundamental Research\\
V.N.Purav Marg, Mumbai-400088,, India.}
}
\begin{document}
\thispagestyle{empty}
\begin{abstract}
The $\tau$ polarization can be easily measured at LHC in the 1-prong 
hadronic $\tau$ decay channel by measuring what fraction of the
\taujet momentum
is carried by the charged track. A simple cut requiring this fraction to be 
$>$0.8 retains most of the $\ptau=$+1 {\taujet} signal while suppressing the 
$\ptau=$-1 {\taujet} background and practically eliminating the fake
$\tau$ background. This can be utilized to extract the charged Higgs 
signal. It 
can be also utilized to extract  the SUSY signal in the stau NLSP region
, and in particular the stau co-annihilaton region.
\end{abstract}
\maketitle

\section{Introduction}
It is easy to measure $\tau$ {\pol}$\ptau$ as it is reflected in the kinematic
distribution of its decay products. Moreover, the best channel for measuring
$\tau$ \pol is also the best channel for $\tau$ identification, i.e. the 
1-prong
hadronic $\tau$ decay channel. In particular a simple kinematic cut, requiring
the single charged prong to carry $>$ 80\% of the hadronic \taujet ~momentum
retains most of the $\ptau$=+1 \taujet events, while suppressing the 
$\ptau$=-1
\taujet background and practically eliminating the fake $\tau$ background 
from standard
hadronic jets. Interestingly the most important channel for charged 
Higgs boson search at LHC is its $\tau$ decay channel, 
$H^- \to \tau_R^{-}{\bar \nu_R}$, giving $\ptau=$+1. Similarly a very
important part of the parameter space of the minimal supergravity(mSUGRA)
model has $\tilde B$ as the lightest superparticle, while the next to the
lightest one is a stau($\stau1$) with a dominant $\tilde\tau_R$ component.
In this case one expects the supersymmetric(SUSY) signal at LHC to 
contain a $\ptau=+1$
$\tau$ from the cascade decay of squarks and gluinos via 
$\stau1 \to \tau_R^-\tilde B$. In both cases one can use the above
kinematic cut to enhance the $\ptau=+1$ signal over the $\ptau=-1$ 
background as well as the fake $\tau$ background.

    The paper is organised as follows. 
In section 2 we summarise the formalism of $\tau$ \pol in the 1-prong
hadronic decay channel and discuss how the abovementioned kinematic
cut retains most of the detectable $\ptau=+1$ $\tau$-jet signal while
supressing the $\ptau$=-1 \taujet as well as the fake \taujet backgrounds.
Section 3 briefly introduces the SUSY search programme at LHC via SUSY 
as well as SUSY Higgs(and in particular $\ch$) signals.
In section 4, we describe the most important $H^\pm$ signal
in both $m_{H^\pm} <m_t$ and  $m_{H^\pm} > m_t$ regions, which contains
a hard $\tau$ with $\ptau=+1$ from the abovementioned $H^\pm$ decay.
In section 5 we show 
Monte Carlo simulations using the above kinematic cut for extraction of the 
$\ch$ signal at LHC for both the  $m_{H^\pm} <m_t$ and  $m_{H^\pm} > m_t$ 
regions. In the latter case we also briefly discuss a corresponding 
kinematic cut for extracting the $m_{H^\pm}$ signal in the 3-prong
hadronic decay channel of $\tau$. 
In section 6 we briefly describe
the SUSY signal coming from the abovementioned cascade decay propcess. We
also emphasize a very important part of the SUSY parameter space, called
the stau co-annihilation region, where the signal contains a soft $\tau$
with $\ptau=+1$. 
In section 7 we show the use of the 
kinematic cut for extracting the SUSY signal at LHC in the 1-prong hadronic 
$\tau$-decay channel, with particular emaphsis on the stau co-annihilation 
region.

\section*{2.~$\tau$ Polarization:}
   The best channel for $\tau$- polarization is its 1-prong hadronic
decay channel, accounting for 50\% of its decay width. Over 90\% this comes
from 
\be
\tau \to \pi^\pm\nu(12.5\%), \rho^\pm \nu(26\%), a_1^\pm \nu(7.5\%),
\label{eq:twentytwo}
\ee
where the branching fraction for $\pi$ and $\rho$ include the small K and 
K$^\ast$ contributions, which have identical polarization effects~\cite{pdg}.
The CM angular distributions of $\tau$ decay into $\pi$ and vector meson 
$v$(=$\rho,a_1$) is simply given in terms of its polarization as 
\br
{1 \over \Gamma_\pi} {d\Gamma_\pi \over d\cos\theta} =
{1\over2} (1 + P_\tau \cos\theta)  \nonumber
\er
\be
{1 \over \Gamma_v} {d\Gamma_{v L,T} \over d\cos\theta} =
{{1\over2} m^2_\tau, m^2_v \over m^2_\tau + 2m^2_v} (1 \pm P_\tau \cos\theta)
\label{eq:twentythree}
\ee
where L,T denote the longitudinal and transverse polarization states of the
vector meson. The fraction $x$ of the $\tau$ laboratory momentum carried
by its decay meson, i.e the(visible) $\tau$-jet, is related to the angle 
$\theta$ via
\be
x = {1\over2} (1 + \cos\theta) + {m^2_{\pi,v} \over 2m^2_\tau} (1 -
\cos\theta),
\label{eq:twentyfour}
\ee
in the collinear approximation($p_\tau\gg m_\tau$). It is clear 
from eqs.~\ref{eq:twentythree} and~\ref{eq:twentyfour} that the 
relatively hard part of the
signal($P_\tau$=+1) $\tau$-jet comes from the $\pi,\rho_L$ and $a_{1L}$ 
contributions, while for the background($\ptau$=-1) $\tau$-jet it comes 
from the $\rho_T$
and $a_{1T}$ contributions\cite{bullock}. Note that this is the important 
part that would pass the $p_T$ threshold for detecting $\tau$-jets.

                     One can simply understand the above feature from angular
momentum conservation. For $\tau_{R(L)} \to \nu_L \pi^-, v^-_{\lambda=0}$
it favors forward(backward) emission of $\pi$ or longitudinal vector meson,
while it is the other way around for transverse vector meson emission, 
$\tau^-_{R(L)} \to \nu_L v^-_{\lambda=-1}$. After boosting back to the 
laboratory frame the forward emitted meson becomes the leading particle, 
giving a hard $\tau$-jet.

            Now the $\rho_T$ and $a_{1T}$ decays favor equal sharing of the
momentum among the decay pions, while the $\rho_L$ and $a_{1L}$ decays 
favor unequal sharing, where the charged pion 
carries either very little or most of the $\tau$-jet momentum. Thus plotted
as a function of the momentum fraction carried by the charged pion, 
\be
R = \frac{p_{\pi^\pm}}{p_{\tau-jet}},
\label{eq:twentyfive}
\ee
the longitudinal $\rho$ and $a_1$ contributions peak at very low of high
R($\lsim$ 0.2 or $\gsim$0.8), while the transverse contributions peak 
in the middle~\cite{bullock,sreerup}. This is shown in Fig. 1~\cite{sreerup}.
\begin{figure}[hbt]
\includegraphics[width=7.5cm, height=5cm]{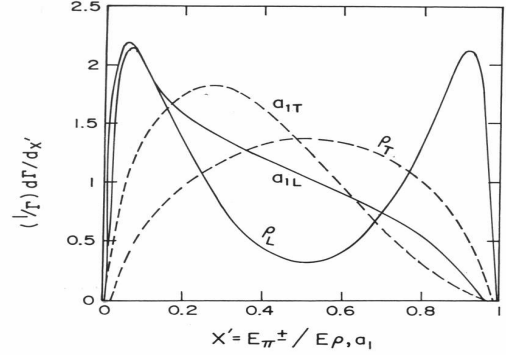}
\vspace{-1.0cm}
\caption{Distributions of $\tau \to \rho \nu, a_1 \nu$ events in the 
fractional hadron($\tau$-jet) energy-momentum carried by the charged
prong X$^\prime \equiv$R[\cite{sreerup},\cite{dp1}].Note that 
$\tau^\pm \to \pi^\pm\nu$ contribution corresponds to a delta function
at X$^\prime =$1  
}
\label{fig:figone}
\end{figure}
Note that the $\tau^\pm \to \pi^\pm \nu$ contribution would appear 
as a delta function at R=1 in this figure. The low R peaks of the 
longitudinal $\rho$ and $a_1$ contributions are not detectable because
of the minimum $p_T$ requirement on the charged track for $\tau$-identification
(R$\gsim$0.2). Now moving the R cut from 0.2 to 0.8 cuts out the
transverse $\rho$ and $a_1$ peaks, while retaining the detectable longitudinal
peak along with the single $\pi^\pm$ contribution. Thanks to the
complimentarity of these two sets of contributions, one can effectively
suppress the former while retaining most of the latter by a simple cut on
the ratio
\be
R >0.8.
\label{eq:twentysix}
\ee
Thus one can suppress the hard part of the $\tau$ jet backround($\ptau$=-1)
while retaining most of it for the detectable signal($\ptau=+$1), even 
without separating
the different meson contributions from one another~\cite{sreerup}. This is
a simple but very powerful result particularly for hadron
colliders, where one cannot isolate the different meson contributions to the
$\tau$-jet in (\ref{eq:twentytwo}). The result holds equally well for a more
exact simulation of the $\tau$-jet including the nonresonant contributions.
It should be noted here that the simple \pol cut(\ref{eq:twentysix}) suppresses
not only the $p_\tau=-1$ $\tau$-jet background, but also the fake $\tau$-jet
background from common hadronic jets. This is particularly important for
$\tau$-jets with low $p_T$ threshold of 15-20~GeV, as we shall need for
SUSY search in the stau co-annihilation region in section 7. Imposing this
cut reduces the faking efficiency of hadronic jets from 5-10\% level to
about 0.2\%. The reason is that a common hadronic jet can fake an 1-prong
$\tau$-jet by a rare fluctuation, when all but one of the constituent
particles(mostly pions) are neutral. Then requiring the single charged
particle to carry more than 80\% of the total jet energy requires a second
fluctuation, which is even rarer.

\section*{3.~SUSY and SUSY Higgs searches at LHC} 
The minimal suupersymmetric standard model(MSSM), has been the most popular
extension of the standard model(SM) for four reasons. It provides (1) a natural
solution to the hierarchy problem of the electroweak symmetry breaking(EWSB)
 scale of the SM, (2) a natural(radiative) mechanism for EWSB,(3) a natural 
candidate for the dark matter of the universe in terms of the lightest 
superparticle(LSP), and (4) unification of the gauge couplings at the grand
unification(GUT) scale. Therefore, there is a great deal of current
interest in probing this model at LHC. This is based on a two-prong search
strategy. On the one hand we are looking for the signal of supersymmetric
(SUSY) particle production at LHC. On the other hand we are also looking for 
the signal of the extended Higgs bosn sector of the MSSM, and in particular 
the charged Higgs boson($H^\pm$). We shall see below that the $\tau$ channel
plays a very important role for both SUSY and the $H^\pm$ signals and one can
use the abovementioned $\tau$ \pol effect in extracting both these signals at 
LHC.   

\section*{4.~$H^\pm$ Signal}
As mentioned above, the MSSM contains two Higgs doublets $H_u$
and $H_d$, the ratio of whose vevs is denoted by $\tan\beta$. The two
complex doublets correspond to 8 degrees of freedom, 3 of which are absorbed
as Goldstone bosons to give masses and longitudinal components to the
${\rm W}^\pm$ and Z bosons. This leaves 5 physical states: two neutral
scalars $h$ and $H$, a pseudo scalar $A$ and a pair of charged 
Higgs bosons
\be
H^\pm = H_u^\pm \cos\beta + H_d^\pm\sin\beta.
\label{eq:fourteen}
\ee
While it may be hard to distinguish any of these neutral Higgs bosons
from that of the SM, the $H^\pm$ pair carry the distinctive
hallmark of the MSSM. Hence the $H^\pm$ search plays a very important
role in probing the SUSY Higgs sector~\cite{dp1}. All the tree level masses
and couplings of the MSSM Higgs bosons are given in terms of $\tan\beta$
and any one of their masses, usually taken to be $m_A$. It is simply
related to $m_{\ch}$ via,
\be
m^2_{\ch} = m_A^2 + m_W^2.
\label{eq:fifteen}
\ee
The most important $\ch$ couplings are
\br
{\ch} t b(cs): \frac{g}{\sqrt{2}M_W} ( m_{t(c)}\cot\beta + m_{b(s)}\tan\beta),
\nonumber
\er 
\be
\ch \tau \nu : \frac{g}{\sqrt{2}M_W}m_\tau\tan\beta.
\label{eq:sixteen}
\ee
Assuming the $H^\pm t b $ coupling to remain perturbative
up to the GUT scale implies $1<\tan\beta<m_t/m_b$.

For $m_{\ch} < m_t$, eq.\ref{eq:sixteen} imply large branching fractions for
\be
t \to b H^\pm
\label{eq:seventeen}
\ee
decay at the two ends of the above range, $\tan\beta \sim 1$ and
$\tan\beta \sim m_t/m_b \simeq$50, driven by the $m_t$ and $m_b$ terms
respectively. But there is a huge dip in the intermediate 
region around
\be
\tan\beta \sim \sqrt{m_t/m_b} \sim 7,
\label{eq:eighteen}
\ee
which is overwhelmed by the SM decay $t\to b W$. Eq.~\ref{eq:sixteen} also
implies that the dominant decay mode for this $H^\pm$ over the theoretically
favored region of $\tan\beta >$1 is,
\be
H^- \to \tau^- \bar\nu_R;~~{\ptau}= +1
\label{eq:nineteen}
\ee
 where the polarization follows simply from angular momentum conservation,
requiring the $\tau^-$ to be right handed. It implies the opposite
polarization for the SM process
\be
W^- \to \tau^- \bar\nu_R;~~\ptau = -1
\label{eq:twenty}
\ee
since the $\tau^-$ is now required to be left-handed. One can use the
opposite polarizations to distinguish the $\ch$ signal from the SM background
~\cite{bullock,sreerup}. In particular one can use the kinematic cut,
mentioned in the introduction, to enhance the signal/background ratio and
extend the $H^\pm$ search at LHC over the intermediate $\tan\beta$
range (\ref{eq:eighteen}), which would not be possible otherwise\cite{sreerup}.

For $m_{\ch} >m_t$, the dominant production process at LHC is the LO
process
\be
gb \to t H^- + h.c.~.
\label{eq:twentyone}
\ee
The dominant decay channel is $H^- \to \bar t b$, which has unfortunately
a very large QCD background. By far the most viable signal comes
from the second largest decay channel(\ref{eq:nineteen}), which has
a branching fraction of $\gsim$10\% in the moderate to large
$\tan\beta(\gsim$10) region. The largest background comes from $t \bar t$
production, followed by the decay of one of the top quarks into the
SM channel(\ref{eq:twenty}). One can again exploit the opposite $\tau$
polarizations to enhance the signal/background ratio and extend the
$\ch$ search to several hundreds of GeV for $\tan\beta \gsim$ 10
\cite{dp2,guchait,les}. This will be discussed in detail in the
next section.

\section*{5.~$\tau$ polarization in the $H^\pm$ search}

A parton level Mone Carlo simulation of the $H^\pm$ signal in the
$m_{H^\pm}<m_t$ region~\cite{sreerup} showed that using the polarization
cut(\ref{eq:twentysix}) enhances the signal/background ration substantially and
makes it possible to extend the $H^\pm$ search at LHC over most of the
intermediate $\tan\beta$ region(\ref{eq:eighteen}). This has been confirmed
now by more exact simulations with particle level event generators.
Fig.2 shows the $H^\pm$ discovery contours at LHC using this polarization
cut\cite{les}. The vertical contour on left shows $H^\pm$ discovery
contour via $t \to b H$ decay. The mild dip 
in the middle shows the remaining gap in this intermediate
$\tan\beta$ region.
\begin{figure}[hbt]
\includegraphics[width=8.0cm, height=5.5cm]{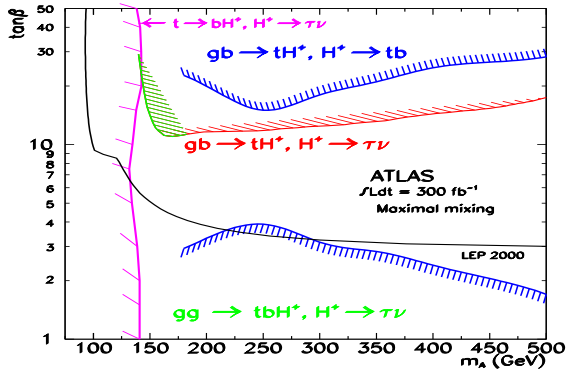}
\vspace{-1.0cm}
\caption{
The 5$\sigma$ $H^\pm$ boson discovery contours of the ATLAS 
experiment at LHC from $t \to b H^\pm, H^\pm \to \tau^\pm \nu$(vertical);
$gb \to t H^-; H^- \to \tau^- \nu$(middle horizontal) and 
$gb \to t H^-; H^- \to \bar t b$ channel\cite{les}.  
}
\label{fig:figone}
\end{figure}

For $m_{H^\pm}>m_t$, the signal comes from
(\ref{eq:twentyone}) and (\ref{eq:nineteen}), while the background comes
from $t \bar t$ production, followed by the decay of one top 
into(\ref{eq:twenty}).
To start with the background is over two orders of magnitude larger than
the signal; but the signal has a harder $\tau$-jet. Thus a
$p_T^{\tau-jet}>$100~GeV cut improves the signal/background ratio. 
\begin{figure}[hbt]
\includegraphics[width=8.5cm, height=7cm]{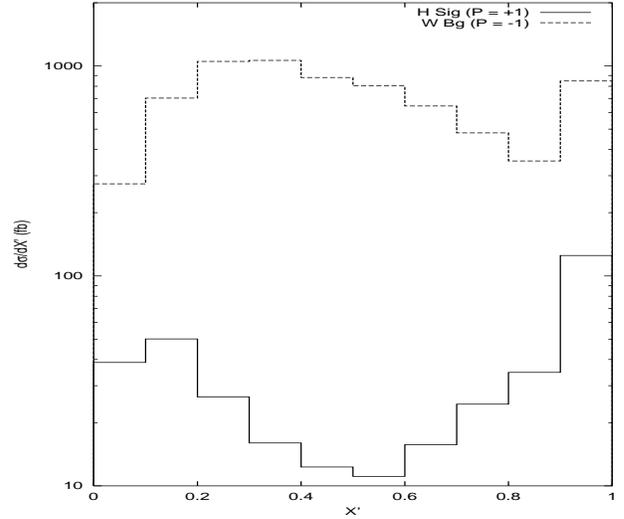}
\vspace{-1.0cm}
\caption{The LHC cross section for a 300~GeV $H^\pm$ signal at 
$\tan\beta=$40 shown along with the $t \bar t$ background in
the 1-prong $\tau$-jet channel, as function of the $\tau$-jet 
momentum fraction X$^\prime$(R) carried by the charged 
pion~\cite{dp1}.
}
\label{fig:figone}
\end{figure}
Fig.3
shows the R(X$^\prime$) distribution of the resulting signal and 
background. One
can see that increasing the R cut from 0.2 to 0.8 suppresses the background
substantially while retaining most of the detectable(R$>$0.2) signal events.
The remaining signal and background can be separated by looking at their
distributions in the transverse mass of the $\tau$-jet
with the missing $p_T$, coming from the accompanying $\nu$. Fig.4
shows these distributions from a recent simulation\cite{guchait} using
{\tt PYTHIA} Monte carlo event generation\cite{pythia}, interfaced with
{\tt TAUOLA}\cite{tauola} for handling $\tau$ decay. One can clearly
separate the $H^{\pm}$ signal from the W background and also measure
the $H^{\pm}$ mass using this plot.
\begin{figure}[hbt]
\includegraphics[width=8.0cm, height=7cm]{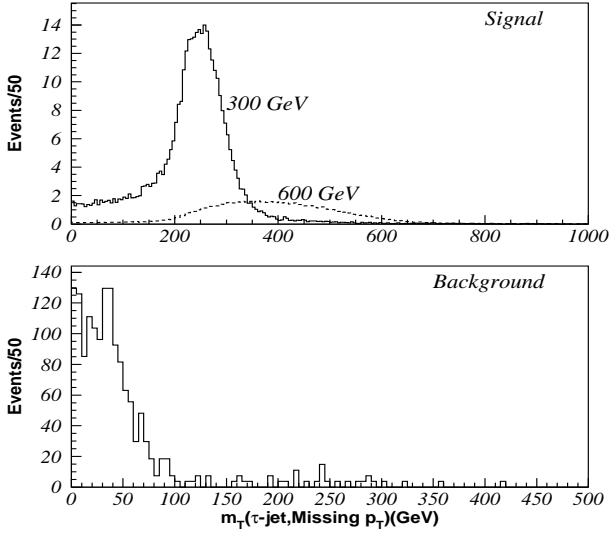}
\vspace{-0.5cm}
\caption{
The number of events are shown against transverse mass $m_T$
for signal and background for 1-prong decay channel of
$\tau$-jets. These are subject to
$p_T^{\tau-jet}> $100 GeV, $R>$0.8 and $E{\!\!\!/}_T >$ 100 GeV.
The masses of charged Higgs are 300~GeV and 600~GeV and $\tan\beta$=40. 
\cite{guchait}.}
\label{fig:figone}
\end{figure}

                 Finally, one can also use the \pol effect in the 3-prong
hadronic $\tau$-decay channel,
\be
\tau^\pm \to \pi^\pm\pi^\pm \pi^\mp \nu,
\label{eq:twentyseven}
\ee
with no neutrals. This has a branching fraction of 10\%, which acounts for
2/3rd of inclusive 3-prong $\tau$-decay (including neutrals). Excluding
neutrals effectively eliminates the fake $\tau$-jet background from common
hadronic jets. About 3/4 of the branching fraction for eq.~\ref{eq:twentyseven}
comes from $a_1$. The momentum fraction $R$ of
$
\pi^\pm\pi^0\pi^0
$
channel is equivalent to the momentum fraction carried by the unlike sign
pion in $a_1 \to \pi^\pm \pi^\pm \pi^\mp$ channel. Thus one sees from Fig.1
that one can retain the $a_{1{\rm L}}$ peak while suppressing $a_{1T}$ by
restricting  this momentum fraction to $<$0.2, which is accessible in this
case. This will suppress the hard $\tau$-jet background events 
from $P_\tau=-1$ while
retaining them for the $\ptau=+1$ signal. This simple result holds even after
the inclusion of the non-resonant contribution. 
\begin{figure}[hbt]
\includegraphics[width=7.5cm, height=5cm]{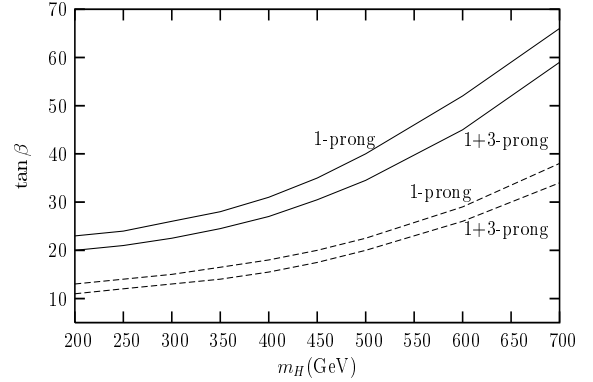}
\vspace{-1.0cm}
\caption{The 5$\sigma$ $H^\pm$ discovery contours at LHC shown for
integrated  luminosities of 30$\lumi$(solid) and 100$\lumi$(dashed)
for $H^\pm \to \tau \nu$, with 1 and 3 prong hadronic decay of 
$\tau$~\cite{guchait}. 
}
\label{fig:figone}
\end{figure}

Fig.5 shows the $H^{\pm}$
discovery contours of LHC using 1-prong and (1+3)-prong
channels~\cite{guchait}. One sees a modest improvement of the discovery
reach by including the 3-prong channel. Note also that the 1-prong
$H^{\pm}$ discovery contour for 100$\lumi$ luminosity is consistent with
that of Fig.2 for the ultimate 300$\lumi$ luminosity of LHC.

\section*{6.~SUSY signal}
We shall concentrate 
in the mSUGRA model as a simple and well-motivated
parametrization of the MSSM. This is described by four and half parameters
~\cite{susy},
\be
m_0, m_{1/2}, A_0, \tan\beta~\rm {and}~sgn(\mu)
\label{eq:one}
\ee
the first three representing the common gaugino and scalar masses and trilinear
coupling at the GUT scale. The $\tan\beta$ stands for the ratio
of the two Higgs vacuum expectation values. The last one denotes the sign
of the higgsino mass parameter whose magnitude is fixed by the radiative
EWSB condition described below. All the weak scale superparticle masses are 
given in terms of these parameters by the renormalization group evolution(RGE).
In particular the gaugino masses evolve like the corresponding gauge
couplings. Thus 
\begin{eqnarray}
M_1 = (\alpha_1/\alpha_G)m_{1/2} \simeq 0.4m_{1/2}, \nonumber 
\end{eqnarray}
\be 
M_2=  (\alpha_2/\alpha_G)m_{1/2} \simeq 0.8 m_{1/2},
\label{eq:two}
\ee 
represent the bino $\tilde B$ and wino $\tilde W_3$ masses 
respectively. A very important weak scale scalar mass, appearing
in the radiative EWSB condition, is 
\be
\mu^2 + M_Z^2 /2 = \frac{M_{H_d}^2 - M_{H_u}^2\tan^2\beta}{\tan^2\beta-1}
\simeq - M_{H_u}^2,
\label{eq:three}
\ee 
where the last equality holds at $\tan\beta>$ 5, favored by the Higgs 
mass limit from LEP~\cite{pdg}. The sign of $M_{H_u}^2$ turning negative by RGE
triggers EWSB, as required by(~\ref{eq:three}). The RHS is related 
to the GUT scale parameters by the RGE,
\begin{eqnarray}
- M_{H_u^2}& = & C_1(\alpha_i, y_t, \tan\beta)m_0^2 + 
C_2(\alpha_i, y_t,\tan\beta)m_{1/2}^2  \nonumber \\ 
& \simeq & -\epsilon ~m_0^2 + 2m_{1/2}^2.
\label{eq:four}
\end{eqnarray}
The tiny co-efficient of $m_0^2$ results from an almost exact cancellation 
of the GUT scale value by a negative top yukawa($y_t$) contribution. We see
from eq.(~\ref{eq:two}-\ref{eq:four}) that apart from a narrow
strip of $m_0 >>m_{1/2}$, the mSUGRA parameter space satisfies the mass
hierarchy 
\be
M_1 < M_2 <\mu.
\label{eq:five}
\ee
Thus the lighter neutralinos and chargino are dominated by the gaugino 
components 
\be
\N0_1 \simeq \tilde B; ~~~~~\N0_2, \Cpm_1 \simeq \tilde W_3, 
\label{eq:six} 
\ee
while the heavier ones are dominated by the higgsino. The lightest neutralino
$\N0_1$($\equiv \tilde\chi$) is the LSP.
 
The ligtest sfermions are the right-handed sleptons, getting only the U(1)
gauge contributions to the RGE, i.e 
\be
m_{\tilde\ell_R}^2 \simeq m_0^2 + 0.15 m_{1/2}^2. 
\label{eq:seven}
\ee
The Yukawa coupling contribution drives the $\tilde\tau_R$ mass still lower. 
Moreover, the mixing between the $\tilde\tau_{L,R}$ states, represented 
by the off-diagonal term,
\be
m_{LR}^2 = m_\tau (A_\tau + \mu \tan\beta),
\label{eq:eight} 
\ee 
drives the lighter mass eigenvalues further down.
Thus the lighter stau mass eigenstate,
\be
\stau1 = \tilde\tau_R \sin\theta_{\tilde\tau} + 
\tilde\tau_L \cos\theta_{\tilde\tau},
\label{eq:nine}
\ee
is predicted to be the lightest sfermion. Moreover, one sees from 
eq.~\ref{eq:two},~\ref{eq:five} and \ref{eq:seven} 
that $\stau1$ is predicted to be the
next to lightest superparticle(NLSP) over half of the parameter space 
\be
m_0 < m_{1/2}.
\label{eq:ten}
\ee 
Thanks to the modest $\tilde\tau_L$ component in eq.(~\ref{eq:nine}),
a large part of the SUSY cascade decay signal at LHC proceeds via
\be
\Cpm_1 \to \stau1\nu \to \tau\nu\N0_1,
\label{eq:eleven} 
\ee
\be
\N0_2 \to \tau'\stau1 \to \tau' \tau \N0_1.
\label{eq:tweleve}
\ee
The dominance of the $\tilde\tau_R$ component in $\stau1$ implies that
the {\pol}
\be
\ptau \simeq +1,
\label{eq:thirteen}
\ee
while $P_{\tau'} \simeq $-1. We shall see in the next section that the {\pol} 
effect can be utilized to extract the SUSY signal containing a 
positively polarized $\tau$ from eq.(~\ref{eq:eleven},\ref{eq:tweleve}).

A very important part of the abovementioned parameter space is the
stau co-annihilation region~\cite{spa}, where the $\tilde B$ LSP
co-annihilates with a nearly degenerate $\stau1$, 
$\N0_1 \stau1 \to \tau \gamma$, to give a cosmologically compatible
relic density~\cite{wmap}. The mass degeneracy $m_{\stau1} \simeq m_{\N0_1}$
is required to hold to $\sim$5\%, since the freeze out temeparture 
is $\sim$5\% of the LSP mass. Because of this mass denegenercay the 
positively polarized $\tau$ lepton coming from 
eqs.(~\ref{eq:eleven},\ref{eq:tweleve}) is rather soft.
We shall see in the next section how the \pol effect can be 
utilized to extract the soft $\tau$ signal and also to measure the tiny mass
difference between the co-annihilating particles.

\section*{7.~$\tau$ \pol in SUSY search:}   
 The \pol of $\tau$ coming from the $\stau1$ decay of eqs.\ref{eq:eleven} and 
\ref{eq:tweleve} is given in the collinear approximation by~\cite{nojiri}
\begin{eqnarray}
P_\tau = \frac{\Gamma(\tau_R) - \Gamma(\tau_L)}{\Gamma(\tau_R)+
\Gamma(\tau_L)}
= \frac{(a^R_{11})^2 - (a^L_{11})^2}{(a^R_{11})^2 + (a^L_{11})^2}
\label{eq:ppol}
\end{eqnarray}
\br
a^R_{11} &=& - \frac{2g}{\sqrt{2}}N_{11}\tan\theta_W \sin\theta_{\tilde\tau}
\nonumber \\
 &-& \frac{g m_\tau}{\sqrt{2} m_W \cos\beta}N_{13}\cos\theta_{\tilde\tau}
\er
\br
a^L_{11} &=& \frac{g}{\sqrt{2}}[ N_{12} + N_{11} \tan\theta_W]
\cos\theta_{\tilde\tau}
\nonumber \\ 
&-&\frac{g m_\tau}{\sqrt{2} m_W \cos\beta} \sin\theta_{\tilde\tau}N_{13}
\label{eq:twentyeight}
\er
where the 1st and 2nd subscript of $a_{ij}$ refer to $\tilde\tau_i$ and
$\N0_{j}$; and
\begin{eqnarray}
\tilde\chi \equiv \N0_1 = N_{11}\tilde B + N_{12}\tilde W_3 +
N_{13}\tilde H_d + N_{14}\tilde H_u
\label{eq:twentynine}
\end{eqnarray}
 gives the composition of LSP. Thus the dominant term is 
 $a_{11}^R \simeq - \frac{2g}{\sqrt{2}}N_{11}\tan\theta_W\sin\theta_
{\tilde \tau}$, implying $\ptau\ \equiv +1$. In fact in the mSUGRA model 
there  is a cancellation between the subdominant terms, so that one 
gets $\ptau>$0.9 throughtout the allowed parameter space\cite{guchait2}. 
Moreover, in the $\stau1$ NLSP region of eq.(\ref{eq:ten}) 
$\ptau >$0.95, so that one can approximate it to $\ptau=+1$. The \pol
of the $\tau'$ from eq.(\ref{eq:tweleve}) is obtained from 
eq.(\ref{eq:ppol})
by replacing $a_{11}^{L,R}$ by $a_{1,2}^{L,R}$. The dominant contribution 
comes from  
$a_{12}^L \simeq \frac{g}{\sqrt{2}}N_{22}\cos\theta_{\tilde \tau}$,
implying $P_{\tau'} \equiv-1$. There is a similar cancellation of the
subdominant contributions, leading to $P_{\tau'}<-0.95$ in the $\stau1$ 
NLSP region. Thus one can safely approximate $P_{\tau'} = -1$.   
\begin{figure}[hbt]
\includegraphics[width=7.5cm, height=5cm]{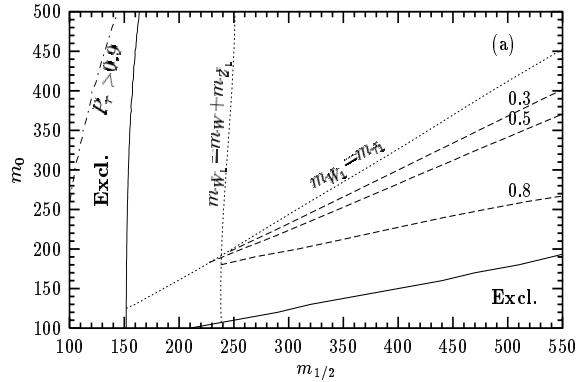}
\vspace{-0.5cm}
\caption{
$BR(\tilde W_1 \rightarrow \tilde\tau_1 \nu_\tau)$ is shown
as contour plots(dashed lines) in $m_0$ and $m_{1/2}$ plane for
$A_0$ = 0, $\tan\beta$ = 30 and
positive $\mu$. The kinematic
boundaries(dotted lines) are shown for $\tilde W_1 \rightarrow W \tilde
Z_1 $ and $\tilde W_1 \rightarrow \tilde\tau_1 \nu_\tau$ decay.
The entire region to the right of the boundary(dot-dashed line) corresponds
to $P_\tau >$0.9.
The excluded
region on the right is due to the $\tilde\tau_1$ being the LSP while that on
the left is due to the LEP constraint 
$m_{\tilde W_1^\pm} >$102 GeV\cite{guchait2}.
Note that here $\tilde W_1$ and $\tilde Z_1$ correspond to 
$\tilde\chi_1^\pm$ and $\tilde\chi_1^0$ in the text.
}
\label{fig:figone}
\end{figure}
    Fig.6 shows that $P_\tau$ is $>$0.9 for $\stau1 \to \tau \N0$ decay 
throughout the mSUGRA parameter space\cite{guchait2}. It also shows
that the branching fraction of the decay(\ref{eq:eleven}) is large over
the $\stau1$ NLSP region of eq.(\ref{eq:ten}), so that one expects a large 
part of the SUSY signal in the $\MET$ channel to contain a $\tau$-jet with
$\ptau$=+1.
\begin{figure}[hbt]
\includegraphics[width=7.5cm, height=5cm]{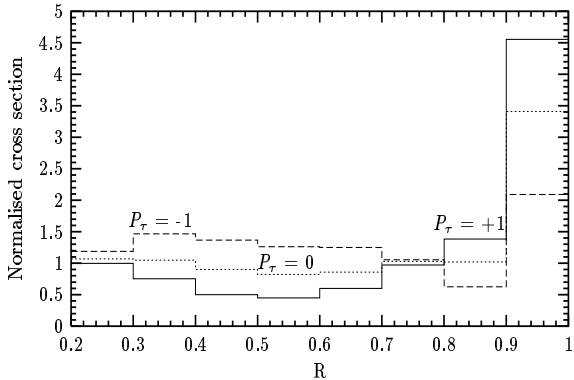}
\vspace{-0.5cm}
\caption{
The normalised SUSY signal cross sections for
$P_\tau$=1(solid line), 0(dotted
lines) and -1( dashed lines) in the 1-prong hadronic
$\tau$-jet channel shown as functions of the $\tau$ jet momentum fraction
(R) carried by the charged prong\cite{guchait2}.
}
\label{fig:figone}
\end{figure}
    Fig.7 shows the R distribution of this $\ptau$=+1 $\tau$-jet at 
LHC\cite{guchait2}. For comparison the R-distributions are also shown
for $\ptau$=0 and -1 for this $\tau$-jet. Thus one can test the SUSY 
model or check the composition of $\stau1(\N0)$ by measuring this 
distribution.

   Let us conclude by briefly discussing the use of $\tau$ \pol in probing 
the stau co-annihilation region at LSP, corresponding to 
$m_{\stau1} \simeq m_{\N0_1}$\cite{guchait3}. This is one of the very 
few regions
of mSUGRA parameter space compatible with the 
cosmological measurement
of the dark matter relic density, and the only one which is also compatible
with the muon magnetic moment anomaly\cite{g2}. It corresponds to a narrow
strip adjacent to the lower boundary of Fig.6, which can be totally covered
at LHC. Therefore, the stau co-annihilation region is a region of special
interest to the SUSY search programme at LHC. In particular one is 
looking for a distinctive signature, which will identify the SUSY 
signal at LHC to this
region and also enable us to measure the tiny mass difference between the
co-annihilating particles, $\Delta M = m_{\stau1} - m_{\N0_1}$. Such a 
distinctive
signature is provided by the presence of a soft ($\ptau=+1$) $\tau$-jet
from the $\stau1 \to \tau \N0_1$ decay of 
eqs.(\ref{eq:eleven},\ref{eq:tweleve}) in
the canonical multijet+$\MET$ SUSY signal. 
\begin{figure}[hbt]
\includegraphics[width=8.5cm, height=7cm]{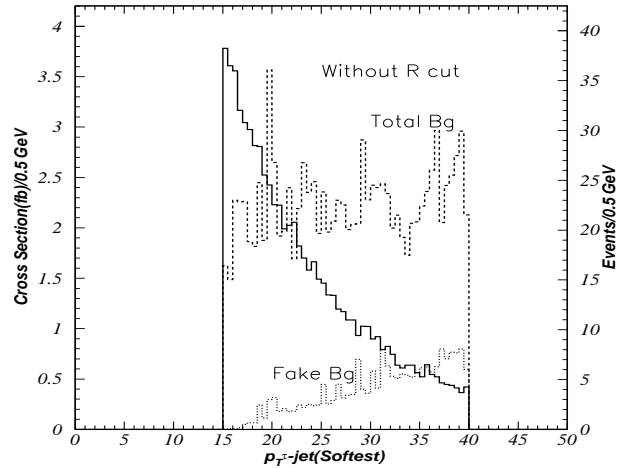}
\vspace{-0.5cm}
\caption{
$p_T$(in GeV) of softest $\tau$-jet for signal and background 
processes\cite{guchait3}.
}
\label{fig:figone}
\end{figure}
\begin{figure}[hbt]
\includegraphics[width=8.5cm, height=7cm]{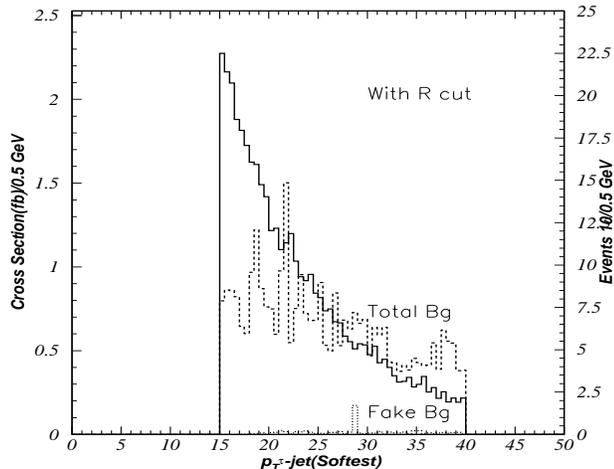}
\vspace{-0.5cm}
\caption{
Same as Fig.8, but with R cut ($>$0.8)~\cite{guchait3}.
}
\label{fig:figone}
\end{figure}
Fig.8 \cite{guchait3} shows the
$p_T$ distributions of this soft ($\ptau=+1)$) $\tau$ jet signal along 
with the 
($\ptau=-1$) $\tau$-jet background coming mainly from the $\N0_2$ decay of 
eq.(\ref{eq:tweleve}) and W decay. It also shows a significant fake $\tau$ 
background from the accompanying hadronic jets in these events. Fig.9
shows that the $R>$0.8 cut of eq.(\ref{eq:twentysix}) effectively
suppresses the ($\ptau=-1$) background to a little over half the signal 
size and 
practically eliminates the fake $\tau$ background. A distinctive signal 
with a very steep slope is clearly sticking above the background at the
low $p_T$ end. One can use this slope to extract the signal from the 
background $\tau$ jets at 3$\sigma$ level with a 10$\lumi$ luminosity
run of LHC, going upto 10$\sigma$ with luminosity of 100$\lumi$. Moreover, 
one can estimate $\Delta M$ to an accuracy of 50\%
at the $\sim$1.5$\sigma$ level with 10$\lumi$, going upto 5$\sigma$
with 100$\lumi$ luminosity\cite{guchait3}.  

\section*{8.~Summary}
The $\tau$ \pol can be easily measured at LHC in its 1-prong hadronic 
decay channel by measuring what fraction of the hadronic $\tau$-jet momentum
is carried by the charged prong. A simple cut requiring this fraction
to be $>$0.8 retains most of the detectable $\ptau=$+1 $\tau$-jet events,
while effectively suppressing the $\ptau=$-1 $\tau$-jet events and 
practically eliminating the fake $\tau$-jet events. We show with 
the help of Monte Carlo simulations that this cut can be effectively used for
(1) Charged Higgs boson and (2) SUSY searches at LHC. (1) The most important
channel for the $H^\pm$ signal at LHC contains a $\ptau=$+1 $\tau$-jet from 
$H^\pm \to \tau \nu$~decay. The above polarization cut can effectively 
suppress the $\ptau=-1$ $\tau$-jet background from $W$ decay, while
retaining most of the detectable signal ($\ptau=+$1) $\tau$-jet events.
So it can be used to extract the $H^\pm$ signal at LHC. (2) Over half the 
mSUGRA parameter space the NLSP is the $\stau1$, which is dominated 
by the right-handed component, while the LSP($\chi$) is dominantly bino. In this
region a large part of the SUSY cascade decay is predicted to proceed
via $\stau1 \to \tau \chi$, giving a $\ptau=+$1 $\tau$-jet along
with the canonical $\MET$+jets. One can use the above polarization cut to 
extract this SUSY signal. A very important part of this region 
is the co-annihilation region, corresponding to $m_{\stau1}\simeq m_{\chi}$.
So the $\ptau=+1$ $\tau$-jet signal is expected to be soft in this region. 
However, one can use this polarization cut to extract this signal from
the $\ptau=-1$ $\tau$-jet and fake $\tau$-jet backgrounds, and also to 
measure the small mass difference between the co-annihilating 
superparticles.

\section*{9.~Acknowledgement}
DPR was supported
in part by the BRNS(DAE) through
Raja Ramanna Fellowship.


\begin{thebibliography}{9}

\bibitem{pdg}
Review of Particle Properties, J. Phys. G33(2006)1.

\bibitem{bullock}
B.K.Bullock, K. Hagiwara and A. D. Martin, Phys. Rev. Lett. 67 (1991) 3055;
Nucl.~Phys. B395 (1993) 499;~D.P.Roy, Phys. Lett. B277 (1992) 183.
\bibitem{sreerup}
S.Raychaudhuri, D.P.Roy, Phys. Rev. D52 (1995) 1556; Phys. Rev. D53(1996)
4902.
\bibitem{dp1}
D.P.Roy Mod. Phys. Lett. A19 (2004) 1813
\bibitem{dp2}
D.P.Roy Phys. Lett. B349 (1999) 607.
\bibitem{guchait}
M.Guchait,R.Kinnunen and D.P.Roy, Euro.Phys.J. C52 (2007) 665.
\bibitem{les}
Higgs working group report(Les Houches 2003):K.A.Assamgon et. al. 
(hep-ph/0406152).
\bibitem{pythia}
T. Sjostrand, P. Eden, C. Friberg, L. Lonnblad, G. Miu,
S. Mrenna and  E. Norrbin, Computer Physics Commun. 135(2001)238.
\bibitem{tauola}
S. Jadach, Z. Was,R. Decker and J. H. Kuehn, Comput.Phys.Commun.76(1993) 361;
P. Golonka {\it et al}, hep-ph/0312240 and references therein.
\bibitem{susy}
For review see, e.g. {\it Perspectives in Supersymmetry,} ed. G.L.Kane, world
scientific(1998); {\it Theory and Phenomenology of sparticles}, M. Drees,
R.M. Godbole and P.Roy; World Scientific(2004);
{\it Weak scale Supersymmetry: From superfields to scattering
events,}, H.~Baer and X.~Tata, Cambridge UK, Univ. Press(2006).
\bibitem{spa}
Supersymmety parameter Analysis; SPA convention and projects, J.A. Aaguilar-
Saavvedra et.al. Euro Phys.J. C46 (2006) 43.
D.P.Roy, AIP Conf. Proc. 939 (2007) 63
(arXiv:0707.1949)[hep-ph].
\bibitem{wmap}
WMAP collaboration, D.N.Spergel, et. al. Astrophys.J.Suppl.148(2003)175,
Astrophy/0302209.
\bibitem{nojiri}
M.M.Nojiri, Phys. Rev.  D51 (1995) 6281; M.M.Nojiri, K. Fujii and
T. Tsukamoto, Phys. Rev. D54 (1996) 6756.
\bibitem{guchait2}
M.Guchait and D.P.Roy, Phys. Rev D54(1996)6756; Phys. Lett. B541(2002)356.
\bibitem{guchait3}
R.M.Godbole, M. Guchait and D.P.Roy, arXiv:0807.2390[hep-ph].
\bibitem{g2}
Muon g-2 collaboration, G. Bennet et. al. Phys. Rev. Lett. 92(2004)
161802.
\end{thebibliography}
\end{document}